# Homogeneity of neutron transmission imaging over a large sensitive area with a four-channel superconducting detector

The Dang Vu[1], Hiroaki Shishido[2,3], Kenji M. Kojima[4,5], Tomio Koyama[5], Kenichi Oikawa[1], Masahide Harada[1], Shigeyuki Miyajima[6], Takayuki Oku[1], Kazuhiko Soyama[1], Kazuya Aizawa[1], Mutsuo Hidaka[7], Soh Y. Suzuki[8], Manobu M. Tanaka[9], Alex Malins[10], Masahiko Machida[10] and Takekazu Ishida[3,5,*]

[1] Materials and Life Science Division, J-PARC Center, Japan Atomic Energy Agency, Tokai, Ibaraki 319-1195, Japan

[2] Department of Physics and Electronics, Graduate School of Engineering, Osaka Prefecture University, Sakai, Osaka 599-8531, Japan

[3] NanoSquare Research Institute, Osaka Prefecture University, Sakai, Osaka 599-8570, Japan

[4] Centre for Molecular and Materials Science, TRIUMF, 4004 Wesbrook Mall, Vancouver, BC V6T 2A3, Canada

[5] Division of Quantum and Radiation Engineering, Osaka Prefecture University, Sakai, Osaka 599-8570, Japan

[6] Advanced ICT Research Institute, National Institute of Information and Communications Technology, 588-2 Iwaoka, Nishi-ku, Kobe, Hyogo 651-2492, Japan

[7] Advanced Industrial Science and Technology, Tsukuba, Ibaraki 305-8568, Japan

[8] Computing Research Center, Applied Research Laboratory, High Energy Accelerator Research Organization (KEK), Tsukuba, Ibaraki 305-0801, Japan

[9] Institute of Particle and Nuclear Studies, High Energy Accelerator Research Organization (KEK), Tsukuba, Ibaraki 305-0801, Japan

[10] Center for Computational Science & e-Systems, Japan Atomic Energy Agency, 178-4-4 Wakashiba, Kashiwa, Chiba 277-0871, Japan

* E-mail: ishida@center.osakafu-u.ac.jp

**Abstract**

We previously proposed a method to detect neutrons by using a current-biased kinetic inductance detector (CB-KID), where neutrons are converted into charged particles using a $^{10}$B conversion layer. The charged particles are detected based on local changes in kinetic inductance of X and Y superconducting meanderlines under a modest DC bias current. The system uses a delay-line method to locate the positions of neutron-$^{10}$B reactions by acquiring the four arrival timestamps of signals that propagate from hot spots created by a passing charged particle to the end electrodes of the meanderlines. Unlike conventional multi-pixel imaging systems, the CB-KID system performs high spatial resolution imaging over a 15 mm×15 mm sensitive area using only four channel readouts. Given the large sensitive area, it is important to check the spatial homogeneity and linearity of detected neutron positions when imaging with CB-KID. To this end we imaged a pattern of $^{10}$B dot absorbers with a precise dot pitch to investigate the spatial homogeneity of the detector. We confirmed the spatial homogeneity of detected dot positions based on the distribution of measured dot pitches across the sensitive area of the detector. We demonstrate potential applications of the system by taking a clear transmission image of tiny metallic screws and nuts and a ladybug. The image was useful for characterizing the ladybug noninvasively. Detection efficiencies were low when the detector was operated at 4 K, so we plan to explore raising the operating temperature towards the critical temperature of the detector as a means to improve counting rates.
(251 words)

Keywords: neutron detector, homogeneity of imaging, linearity, current biased kinetic inductance detector

## 1. Introduction

Neutron beams penetrate deeply into most dense materials and are sensitive to certain light elements such as hydrogen, lithium, boron, carbon and oxygen. These properties are in marked contrast to those of X-ray, electron and proton beams. Neutron imaging has been used as a non-destructive and non-invasive technique to investigate materials by means of Bragg edge transmission spectra [1,2], pore structures in materials [3], material properties [4,5], lithium in solid-state batteries [6,7], water in fuel cells [8] and biological systems [9,10]. The ability to conduct high-spatial resolution neutron imaging is closely related to the development of intense neutron beams, since the neutron flux per pixel decreases as a detector's resolution increases. In recent years, many new neutron sources have come online at nuclear spallation facilities. These intense pulsed neutron beams complement the continuous neutron beams available at research reactors. New high-end neutron facilities are under construction or being planned across the globe, targeting further improvements in beam quality and intensity. In around 20 years, an inertial fusion neutron source is expected to enhance currently available neutron beam intensities by three orders of magnitude [11]. Developments in high resolution neutron detectors will be important for maximizing the imaging potential of high flux density neutron sources.

Our group developed the current-biased kinetic inductance detector (CB-KID) at the Japan Proton Accelerator Research Complex (J-PARC). We demonstrated neutron imaging in two dimensions with four signal readout lines with a spatial resolution of down to 22 µm using a CB-KID with an effective area of $A_d =$ 10 mm×10 mm and a meanderline repetition period of $p = 2$ µm [12]. A spatial resolution of 16 µm was achieved by using a CB-KID with $A_d =$ 15 mm×15 mm and $p = 1.5$ µm [13], however the obtained neutron images were slightly blurred due to the deviation of the neutron beam from parallel because of the finite collimation ratio *L/D* (see below). CB-KID has the merit of only needing a small number of connected signal leads compared to other cryogenic detectors. This is favorable as it minimizes the heat flow from the room to the device appreciably.

We used a delay-line method to perform position-sensitive detection with CB-KID, since the signals propagate towards the ends of the meanderlines at constant velocity. A merit of using a superconducting wave guide in the delay-line system is that signals can propagate for long distances at a constant velocity along the striplines due to low dissipation. This means that a large sensitive area can be achieved with CB-KID. However for practical neutron imaging, a detector needs good spatial homogeneity and linearity. In this study we investigated the spatial homogeneity of CB-KID by imaging a regular test pattern of $^{10}$B dots.

The CB-KID was designed using computer-aided design (CAD) software and fabricated at the Clean Room for Analog-Digital-Superconductivity (CRAVITY) at the National Institute of Advanced Industrial Science and Technology (AIST). Although the fabrication process was supposed to be reproducible enough to build a homogeneous CB-KID device, it remained to be investigated whether spatial homogeneity was good over the full sensitive area. We therefore imaged a test sample containing $^{10}$B dots in a regular array and measured the pitch of the dots across different positions in the sensitive area of the detector. We also imaged several other test samples with CB-KID to identify potential applications of the imaging system and characterized the system detection efficiency.

## 2. Operating principle of the superconducting neutron detector

The operating principle of CB-KID with current-biased superconducting nanowires is fundamentally different to that of other neutron detectors. A phenomenological explanation is as follows. The inductance $L_0$ in a superconducting nanowire is expressed as $L_0 = L_m + L_k = L_m + m_s \ell / n_s q_s^2 S$, where $L_m$ is the magnetic inductance, $L_k$ is the kinetic inductance, $\ell$ is the total wire length, $S$ is the cross-sectional area of the wire, $m_s$ is the effective mass, $q_s$ is the electric charge of the Cooper pair and $n_s$ is the Cooper pair density. A transient change in $n_s$ on the superconducting wire occurs when a local hot spot is created by the partial dissipation of kinetic energy by a charged particle originating from a $^{10}$B-neutron nuclear reaction in a conversion layer (see below). As shown in **figure 1**, when a DC bias current $I_b$ is fed into the superconducting nanowire, a pair of voltage pulses are generated at a hot spot within a tiny portion of the superconducting wire over length $\Delta \ell$ ($\ll \ell$) and the pulses propagate towards each end of the wire as electromagnetic wave packets with opposite polarities. A voltage $V$ across the hot spot is expressed as $V = I_b dL_k/dt = I_b d\Delta L_k/dt$ with a constant DC bias current $I_b$ in the superconducting zero-resistance state and a tiny fractional kinetic inductance $\Delta L_k = m_s \Delta \ell / n_s q_s^2 S$. It is not necessary for the bias current to be close to the critical current of the meanderline. This means that the bias current can be used to tune the signal amplitude so that it fits the sensitivity range of the signal capture instrumentation.

A microscopic explanation of the CB-KID operating principle was given in an earlier work [14]. The CB-KID device has a superconductor-insulator-superconductor (S-I-S) planar waveguide so electromagnetic pulse signals transmit very efficiently as Swihart pulses [15]. The detection mechanism is not related to the Josephson junction because the insulator is thick enough to suppress any Josephson currents. Theoretical arguments for the signal-generation mechanism and its transmission along the meanderline were given by Koyama and Ishida [14]. They proposed a propagation wave equation for a CB-KID signal

$$\frac{1}{v^2}\frac{\partial^2 \theta(x,t)}{\partial t^2} - \frac{\partial^2 \theta(x,t)}{\partial x^2} - \frac{\gamma}{c}\frac{\partial}{\partial t}\frac{\partial^2 \theta(x,t)}{\partial x^2} = F(x,t) \qquad (1)$$

where $\theta(x,t)$ is the superconducting phase of the nanowire, $\gamma$ is a coefficient for quasi-particle damping, $v$ is the propagation velocity and $c$ is the velocity of light in a vacuum. $F(x,t)$ is a source term acting within the hot spot, which stems from the spatiotemporal variation of the order parameter ψ(x,z,t) in the hot spot of the S-I-S waveguide and excites the phase oscillation modes that propagate as Swihart waves along the stripline. They also proposed an interpretation of the empirical equation $V(t) = I_b dL_k/dt$ to explain the emergence of signals using a microscopic expression for $L_k$ from London-Maxwell theory [14]. This theory predicted that the signal polarity depends on the direction of the bias current, i.e. a negative pulse appears at the downstream side of the hot spot while a positive pulse appears at the upstream direction. Note that the kinetic inductance only plays a role in the restricted regime of the hot spot and is independent of the signal transmission at a constant velocity over a long distance along the stripline. Therefore, we expected that CB-KID should have good spatial homogeneity and linearity for imaging, although this needed to be investigated.

## 3. Experiments

The delay-line CB-KID system has four output channels and we image the hot spot distribution on the detector using the arrival timestamps of the pulses at the end of the meanderlines. CB-KID has two orthogonal meanderlines of superconducting Nb nanowires and a superconducting Nb ground plane. Two voltage pulses with opposite polarities originate at a hot spot on a meanderline and propagate as electromagnetic waves towards both ends. A negative pulse propagates in the downstream direction to the DC bias current flow, and a positive pulse propagates in the upstream direction. Using the propagation velocities along the X and Y meanderlines ($v_x$ and $v_y$), and the arrival timestamps $t_{Ch1}$, $t_{Ch2}$, $t_{Ch3}$, and $t_{Ch4}$ of the signals for each channel, the coordinate (X,Y) of the hot spot can be calculated as $X = (t_{Ch4} - t_{Ch3})v_x p/2h$ and $Y = (t_{Ch2} - t_{Ch1})v_y p/2h$. Here, $h$ is the length of each segment and $p$ is the repetition period of the meanderline. The calculated hotspot positions are used to render a neutron transmission image.

The CB-KID was fabricated on a thermally oxidized Si substrate sequentially as (1) a 300 nm thick Nb ground plane, (2) a 350 nm thick SiO$_2$ layer, (3) a 40 nm thick Y meanderline, (4) a 50 nm thick SiO$_2$ layer, (5) a 40 nm thick X meanderline and (6) a 50 nm thick SiO$_2$ layer. On top of the CB-KID, a $^{10}$B neutron conversion layer was created by painting on a mixture solution of GE7031 varnish and $^{10}$B fine particles with a brush and allowing the mixture to dry. The mixture was applied to give a layer thickness that is greater than the range of $^{4}$He and $^{7}$Li ions released from $^{10}$B-neutron reactions. The 0.9 μm wide X and Y meanderlines had 10,000 segment repetitions ($h$ = 15.1 mm). The meanderline repetitions of Nb segments were separated by 0.6 μm wide insulating strips, giving $p$ = 1.5 μm and a total meanderline length of $\ell$ = 151 m.

DC bias currents of $I_b^{(x)} = 100$ μA and $I_b^{(y)} = 100$ μA were fed into the X and Y meanderline of the CB-KID respectively via 3 kΩ bias resistors from two low-noise voltage sources. Note that a bias current of 100 μA remains only 0.6% of the critical current density $J_c$ of a Nb stripline, $5 \times 10^7$ A/cm$^2$ at 4 K. The positive signals from the upstream electrodes of the X (Ch2) and Y (Ch4) meanderlines were amplified by differential ultralow-noise amplifiers (NF Corporation, Model SA-430F5). The negative signals from the downstream electrodes of the X (Ch1) meanderline and Y (Ch3) meanderline were inverted and amplified. The Kalliope-DC circuit received four positive signals with a 1 ns sampling multichannel (16 Ch×2) time-to-digital converter (TDC) [12,16]. The CB-KID was cooled to temperatures below $T_c$ using a Gifford-McMahon cryocooler. The detector temperature and the second-stage temperature of the cryocooler were respectively controlled at 4 K and 3.7 K using temperature controllers (Cryocon Inc., Model 44) with Cernox thermometers and electric heaters.

We determined the signal propagation velocities along the X and Y meanderlines by feeding a pulse signal and measuring the time of travel from one electrode to the other. The velocities are temperature dependent, and were estimated as $v_x = 6.328 \times 10^7$ m/s and $v_y = 5.769 \times 10^7$ m/s at 4 K from a theoretical curve fit to the temperature dependence of measured velocities [14]. During imaging experiments, the CB-KID was controlled at 4 K within a stability of $\pm 4$ mK. Since we used four channels of one TDC unit of the Kalliope-DC circuit to measure the four timestamps $t_{Ch1}$, $t_{Ch2}$, $t_{Ch3}$, and $t_{Ch4}$ using a common clock, the best resolution for measuring a time difference $(t_{Ch4} - t_{Ch3})$ or $(t_{Ch2} - t_{Ch1})$ was 1 ns. Each meanderline has a discrete structure with a repetition period of 1.5 μm for determining $x$ and $y$ coordinates. Thus, the $x$ and $y$ coordinates obtained are not continuous but have discrete coordinate steps for each direction. Average step sizes were obtained as $\Delta x_p = (1\text{ns}) \times v_x p/2h$= 3.1 μm for the $x$ steps and $\Delta y_p = (1\text{ns}) \times v_y p/2h$= 2.8 μm for the $y$ steps. There are some uncertainties in the stability of the propagation velocity, time jitters of signals in the readout circuit, heterogeneities in the striplines of the CB-KID sensor, as well as electromagnetic noise from the environment. Such uncertainties may obscure the property of discreteness to some extent in the coordinates derived from repetitive structure. As the system presumes constant transmission velocities of the signal along the meanderlines, it was necessary to check the local precision and overall spatial linearity of the positions identified by the CB-KID sensor experimentally.

## 4. Results and discussion

4.1. Transmission imaging using the superconducting neutron detector

**Figure 2(a)** shows the test samples used for transmission neutron imaging. Screws (sample #1, sample #4), nuts (sample #2, sample #5), a phosphor plate (sample #3), and a dried insect (sample #6) were mounted on an Al plate. Under the Al plate, we placed an array of $^{10}$B dots manufactured using a stainless-steel mesh (sample #7) (see **figure 2(b)**). The test samples were placed at a distance of 0.8 mm from the CB-KID meanderlines to minimize the broadening effect of the pulsed neutron beam at the MLF facility at J-PARC, where the collimation ratio $L/D$=140 is the collimator length ($L$=14 m) divided by the diameter of the neutron moderator ($D$=0.1 m) [17]. The neutron beam was irradiated from the Si substrate side of the CB-KID.

Due to the 800 μm distance between the samples and detector, the finite value of $L/D$ influences the sharpness of the transmission image on the scale of $\delta$ = 800 μm/($L/D$) = 5.7 μm. However this should not alter distances between two distinct points in a transmission image. This is because while the finite $L/D$ ratio tends to blur images, centroid positions should be basically unchanged [12,13]. The pitch of $^{10}$B dots can be determined by CB-KID with a finite $L/D$ ratio neutron beam. Neutron beams with high flux intensities that are expected to come online in future will facilitate easy-to-use high $L/D$ geometries for transmission imaging.

Heterogeneities in the thickness and density of the $^{10}$B neutron conversion layer mean that neutron intensities are non-uniform across transmission images taken with CB-KID. In a previous work [13] we corrected for these effects by normalizing with intensities of an image taken with no test sample. However, we found that although this method worked partially to remove heterogeneity effects in images, it was not flawless. It was likely that the arrangements of the apparatus were slightly different for the images taken with and without the test sample, due to needing to re-mount the cryogenic detector between measurements. In this study we instead normalized the intensities in transmission images taken using a full range of neutron wavelengths (0.002 nm to 1.13 nm) by dividing by the intensities from an equivalent image taken of the sample using shorter wavelength neutrons (0.002 nm to 0.052 nm) (normalized image in **figure 3**).

The test samples of a stainless-steel screw (sample #1), two Ni-plated brass nuts (samples #2 and #5), a titanium screw (sample #4), a seven-spotted ladybug (insect sample #6), the $^{10}$B dots (sample #7) can be seen clearly in **figure 3**. The 0.1 mm-thick phosphor bronze plate (sample #3) cannot be seen clearly due to weaker contrast. The normalization method worked well because the absorption cross section of $^{10}$B for neutrons is inversely proportional to the neutron velocity, and small for neutrons in the range 0.002 nm and 0.052 nm. An average pixel size of $(5\Delta x_p, 5\Delta y_p) =$ (15.7 µm, 14.4 µm) was chosen to render an image with good contrast for **figure 3**. We utilized pulsed neutrons for $4.5 \times 10^6$ triggers $\simeq 50$ h under a proton beam power of 152 kW, which was a small fraction (15%) of the full power of 1000 kW of the facility. **Figure 3** shows good agreement with the optical photograph of **figure 2(a)**.

4.2. Spatial linearity and homogeneity over the sensitive area of the detector

The $^{10}$B dots were designed to be 100 µm×100 µm squares, but as a consequence of the wet etching fabrication process the manufactured dots had rounded corners (**figure 2(b)**). The size of the holes was reported to be 106 µm by the factory which fabricated the sample. Although transmission images are somewhat smeared due to the finite *L/D* ratio, the effect of this on distances measured between two points should be small. However this effect means that the spatial resolutions reported in earlier papers are larger than the reality for the detector [12,13]. The reproducibility of the hole shape appears to be good over the whole mesh as judged from the photo in **figure 2(b)**, which can be expected as the mold pattern for the stainless-steel plate was fabricated using computer-aided design (CAD). The pitch of the $^{10}$B dots can be obtained with high precision by sensing the dot boundary. In the following the repetition period *p* of the $^{10}$B dots was determined by the difference in edge positions of neighboring $^{10}$B dots.

We denote the pitch along the *X* direction by $P_x$ and along the *Y* direction by $P_y$. In **figure 4(a)**, we show a neutron transmission image of the $^{10}$B-dot pattern embedded in the test sample. To obtain better intensity statistics for analysis, the pixels were binned as $(\Delta x_p, 5\Delta y_p)$ to draw a line profile along the *X* direction and as $(5\Delta x_p, \Delta y_p)$ for the *Y* direction. The line profile of the intensities along the *X* direction had a rectangular-wave shape because of strong absorption of the neutrons by the $^{10}$B dots. The blue dots in **figure 4(b)** show the differential of the central line profile with respect to a position. Positive and negative peaks correspond to the boundaries of the $^{10}$B dots in the stainless-steel mesh. We fitted the positive and negative peaks with Gaussian functions using least-squares to obtain the positions of the dot edges (red solid line of **figure 4(b)**). The regular repetition of $^{10}$B dots can be used as hypothetical tick marks of a ruler across the CB-KID sensitive area. In **figure 4(c)**, we plot the *X* positions of positive peaks as a function of the sequential number of each $^{10}$B dot from left to right for the first row of $^{10}$B dots from the bottom. Data points are fitted by a linear line $x_{\text{dot}} = a\, N_{\text{dot}} + x_0$ with $a = 249.75 \pm 0.11$ µm $x_0 = -7431.8 \pm 3.7$ µm. The gradient $a = 249.75 \pm 0.11$ µm is in good accordance with the designed value for the pitch (=250 µm), noting thermal contraction of the sample might be important at 4 K. This demonstrates that the linearity of CB-KID is good along the $X$ direction. Although the $^{10}$B dots did not extend for long enough across the sample to examine linearity along the $Y$ direction, we reasonably expect that the linearity is similar to the $X$ direction in view of the XY symmetric structure of CB-KID.

We compared estimated pitches to investigate the spatial homogeneity of CB-KID imaging over the sensitive area. Pitches were estimated from the line profiles as distances between two neighboring positive peaks, or two neighboring negative peaks, which represent the distances between the left-hand and right-hand sides of neighboring $^{10}$B dots, respectively (see **figure 3** and **figure 4(a)**). We obtained four histograms of dot pitch estimates, *i.e.* two for the $X$ direction and two for the $Y$ direction. The histograms show sharp peaks at around 250 μm. **Figure 5** shows the distribution of distances between negative-peak positions along the $X$ direction, while **Figure 6** shows the same but along the $Y$ direction. The histograms were fitted with Gaussian functions of the form $F = F_0 + A \exp\{-(D_x - P_x)^2/(H_x^2/4\ln 2)\}$, where $D_x$ (or $D_y$) is the distance between neighboring dot edges along the $X$ (or $Y$) direction and $F_0$ is an offset factor representing the noise floor in the central-line profile (which was very small in practice). $H_x$ (or $H_y$) is the full width at half maximum (FWHM). The obtained peak positions of the Gaussian functions give estimates of the pitch of the $^{10}$B dots $P_x$ (or $P_y$). Since the data processing of the central-line profile for each combination of the neighboring dots was undertaken by eye, the results are subtly different depending on the choice for the central line position. Therefore, we repeated the analysis procedure seven independent times, giving fourteen sets of $P_x, H_x, P_y,$ and $H_y$. The fourteen results were averaged to obtain the pitch and the uncertainty, where the standard deviations were regarded as errors for the estimated parameters. Finally, we concluded that the pitch was $\langle P_x \rangle = 250.70 \pm 0.04$ μm with a scatter of $\langle H_x \rangle = 5.36 \pm 0.16$ μm for the $X$ direction, and the pitch was $\langle P_y \rangle = 249.09 \pm 0.07$ μm with a scatter of $\langle H_y \rangle = 3.37 \pm 0.20$ μm for the $Y$ direction. The full widths at half maxima ($\langle H_x \rangle$ and $\langle H_y \rangle$) are good measures of the spatial variation of the measured pitches. The low spatial variations show that the CB-KID has good homogeneity for imaging the $^{10}$B dots. The results mean that the positions (or distances) obtained from the image are rather homogeneous over the sensitive area of the detector.

Note that this result can be achieved wherever the test entities are located in the sensitive field of view of the detector. The pitch dispersion is smaller for the $Y$ direction than for the $X$ direction because of the difference between velocities $v_x$ and $v_y$ ($\because v_x > v_y$). We argue that CB-KID showed good linearity and homogeneity along the $X$ direction in determining positions in neutron imaging. Since the instrumental limitation is mainly determined by the time resolution (1 ns) of the time-to-digital converter, we believe that further improvements in the local positional determination of the detector will be attainable by employing a time-to-digital converter with a faster sampling clock ($\ll 1$ ns).

### 4.3. Application to entomology

A potential application of neutron imaging is for entomological research. A clear transmission image of the seven-spotted ladybug (sample #6) with an average pixel size of $(5\Delta x_p, 5\Delta y_p)$ is shown in **figure 3**. We labelled the two elytra of the ladybug, which are reported to be thinner than 60 μm [18,19], six legs, the pronotum, eyes, antennas, the flight muscle, and cuticula for supporting the wings. The details visible in the neutron transmission image suggest that the technique could be useful for noninvasive characterization of insects. It is well recognized that sex differentiation of ladybugs is difficult with noninvasive techniques. Although the genitalia supported by the cuticula cannot be seen clearly in the image in **figure 3**, the asymmetric shape of the organ visible in the image is consistent with female genitalia shown in anatomical drawings [20].

4.4. Detection efficiency of the superconducting neutron detector

In **figure 7**, we show the detection efficiency of the Y meanderline at 4 K, where the neutron intensity was experimentally determined as a function of neutron wavelength using of Particle and Heavy Ion Transport code System (PHITS) simulations [21]. The detection efficiency of the X meanderline at 4 K was similar to that of the Y meanderline. The detection efficiency of the simultaneous measurements of the X meanderline and the Y meanderline at 4 K is also shown for comparison. We note that the efficiency of simultaneous detection events by the X and Y meanderlines at 4 K was rather low. Potentially the hot spot size is smaller than the repetition period of the meanderline, or hot spots have low efficiency for producing measurable signals.

The efficiency of simultaneous X and Y detections is much smaller than that estimated from PHITS simulations [22], as is evident from **figure 7**. This is likely because the PHITS simulations did not consider the effects of the detector temperature, the bias current, signal processing and dead-time. The simulations mainly accounted for geometrical effects. We emphasize that an important conclusion from the PHITS simulations was that CB-KID has low sensitivity to gamma rays [22]. This is a desirable property of CB-KID contributing to its usefulness as a neutron detector even under strong gamma rays.

The number of events as a function of wavelength measured by using the Y meanderline (**figure 7**) reproduced well the known profile for neutrons at BL10 of the J-PARC facility [12]. Operating CB-KID at various different temperatures revealed that the number of counting events increased remarkably when the operating temperature increased to near the critical temperature of the meanderline [23]. This suggests that hot-spot sizes become larger than the repetition period of the meanderline at temperatures close to the critical temperature [23]. We are currently considering how to increase the detection efficiency of simultaneous X and Y measurements of the meanderlines. We believe that the temperature-dependent detection efficiency offers a clue for improving the efficiency of CB-KID as a four-channel detector. Higher detection efficiencies may be achieved by operating the detector at higher temperatures.

5. Conclusions

In summary, we demonstrated that a four-readout superconducting neutron imaging system shows good spatial heterogeneity and linearity using the delay-line method. We examined the precise pattern of a $^{10}$B-dot-array absorber and found the *X*-direction pitch $\langle P_x \rangle = 250.7$ μm with a scatter of $\langle H_x \rangle = 5.4$ μm and the *Y*-direction pitch $\langle P_y \rangle = 249.1$ μm with a scatter of $\langle H_y \rangle = 3.4$ μm while the $^{10}$B dot array was fully extended toward the *X* direction across the detector sensitive area. We consider that this demonstrates detection with good spatial homogeneity and conclude that this is because the fabrication process at CRAVITY being precise enough to guarantee the invariance of the signal transmission velocities along the meanderlines. The transmission-imaging system is suitable for use at pulsed neutron facilities. We expect that enhancing the time resolution of the TDC circuit will further improve performance. In addition, we also reported a clear transmission image of tiny screws and nuts, and ladybug (insect). We recognize that further efforts to improve the detection efficiency are necessary to make using the detector more practical.

**Acknowledgements**

This work was partially supported by a Grant-in-Aid for Scientific Research (Grant No. JP16H02450, JP19K03751) from JSPS. The neutron-irradiation experiments at the Materials and Life Science Experimental Facility (MLF) of the J-PARC were conducted under the support of MLF programs (Proposals Nos. 2017A0011, 2017B0014, 2018A0109, 2019A0004, 2018P0201, 2019P0201, 2020P0201). We thank Prof. M. Ishii and Dr. M. Itakura for useful discussions on the properties of the ladybug.

ORCID iDs
The Dang Vu https://orcid.org/0000-0002-2579-4164
Hiroaki Shishido https://orcid.org/0000-0003-2196-6827
Shigeyuki Miyajima https://orcid.org/0000-0002-5153-5537
Alex Malins https://orcid.org/0000-0003-1922-4496
Takekazu Ishida https://orcid.org/0000-0002-9629-5178

**References**

[1] Kardjilov N, Manke I, Hilger A, Strobl M and Banhart J 2011 Mater. Today **14** 248

[2] Steuwer A, Withers P J, Santisteban J R and Edwards L 2005 J. Appl. Phys. **97** 074903

[3] Brooks A J, Ge J, Kirka M M, Dehoff R R, Bilheux H Z, Kardjilov N, Manke I and Butler L G 2017 Prog. Addit. Manuf. **2** 125

[4] Woracek R, Penumadu D, Kardjilov N, Hilger A, Boin A, Banhart J and Manke I 2014 Adv. Mater. **26** 4069

[5] Tremsin A S, Rakovan J, Shinohara T, Kockelmann W, Losko A S and Vogel S C 2017 Scientific Reports **7** 40759

[6] Siegel J B, Stefanopoulou A G, Hagans P, Ding Y and Gorsich D 2013 J. Electrochem. Soc. **160** A1031

[7] Manke I, Banhart J, Haibel A, Rack A, Zabler S, Kardjilov N, Hilger A, Melzer A and Riesemeier H 2007 Appl. Phys. Lett. **90** 6

[8] Satija R, Jacobson D L, Arif M and Werner S A 2004 J. Power Sources **129** 238

[9] Isaksson H, Le Cann S, Perdikouri C, Turunen M J, Kaestner A, Tägil M, Hall S A and Tudisco E 2017 Bone **103** 295

[10] Witzmann F, Scholz H, Mueller J and Kardjilov N 2010 Zool. J. Linn. Soc. **160** 302

[11] Taylor A, Dunne M, ennington S, Ansell S, Gardner I, Norreys P, Broome T, Findlay D and Nelmes R 2007 Science **315** 1092

[12] Shishido H, Miki Y, Yamaguchi H, Iizawa Y, Vu T D, Kojima K M, Koyama T, Oikawa K, Harada M, Miyajima S, Hidaka M, Oku T, Soyama K, Suzuki S Y and Ishida T 2018 Phys. Rev. Appl. **10** 044044

[13] Iizawa Y, Shishido H, Nishimura K, Vu T D, Kojima K M, Koyama T, Oikawa K, Harada M, Miyajima S, Hidaka M, Oku T, Soyama K, Aizawa K, Suzuki S Y and Ishida T 2019 Supercond. Sci. Technol. **32** 125009

[14] Koyama T and Ishida T 2018 J. Phys.: Conf. Ser. **1054** 012055

[15] Swihart J C 1961 J. Appl. Phys. **32** 461

[16] Kojima K M, Murakami T, Takahashi Y, Lee H, Suzuki S Y, Koda A, Yamauchi I, Miyazaki M, Hiraishi M, Okabe H, Takeshita S, Kadono R, Ito T, Higemoto W, Kanda S, Fukao Y, Saito N, Saito M, Ikeno M, Uchida T and Tanaka M M 2014 J. Phys. Conf. Ser. **551** 012063

[17] Oikawa K, Maekawa F, Harada M, Kai T, Meigo S, Kasugai Y, Ooi M, Sakai K, Teshigawara M, Hasegawa S, Futakawa M, Ikeda Y and Watanabe N 2008 Nucl. Instrum. Meth. Phys. Res. A **589** 310

[18] Zohry N M H and El-Sayed A M 2019 J. Basic Appl. Zool. **80** 16

[19] Xiang J, Du J, Li D and Scarpa F 2017 J. Mater. Sci. **52** 13247

[20] Wang X, Tomaszewska W and Ren S 2014 ZooKeys **448** 37

[21] Harada M, Maekawa F, Oikawa K, Meigo S, Takada H and Futakawa M 2011 Nucl. Sci. Technol. **2** 872

[22] Malins A, Machida M, Vu T D, Aizawa K and Ishida T 2020 Nucl. Instrum. Meth. Phys. Res. A **955** 163130

[23] Vu T D, Iizawa Y, Nishimura K, Shishido H, Kojima K M, Oku T, Soyama K, Aizawa K, Koyama T and Ishida T 2019 J. Phys. Conf. Ser. **1293** 012051

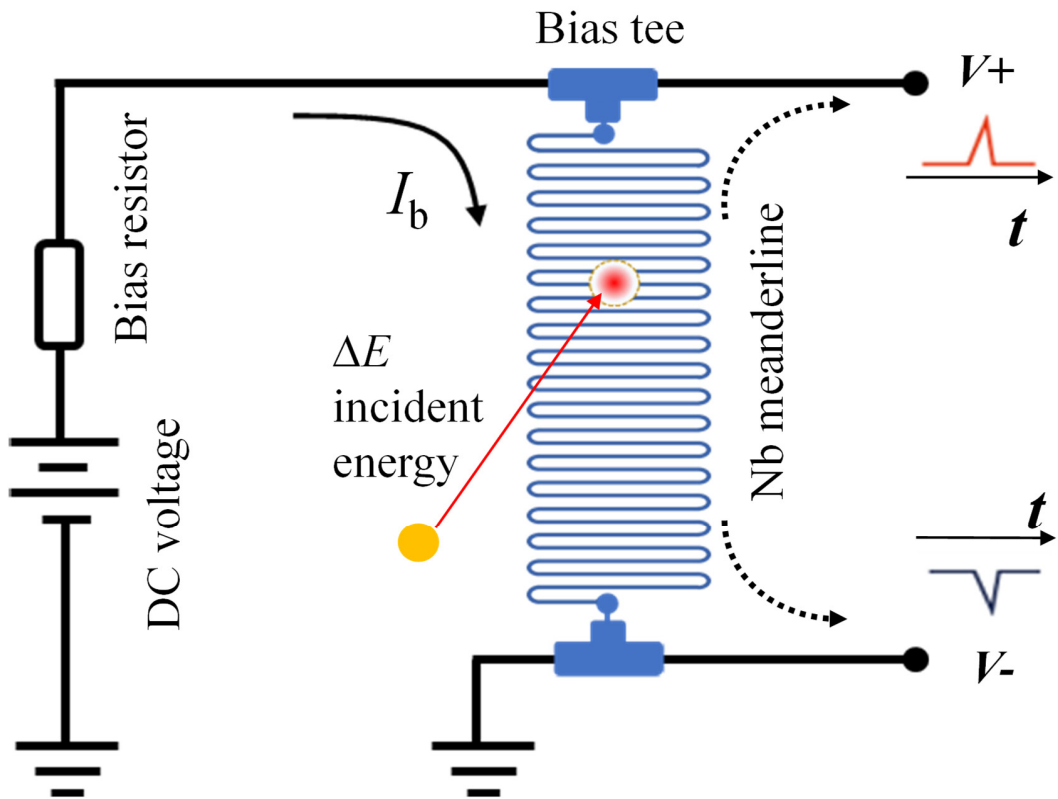

**Figure 1.** Schematic structure of a single CB-KID meanderline fed by a small DC bias current. When a hot spot (red circle) appears in the Nb meanderline, a positive signal propagates towards the upstream side of the DC bias current and a negative signal propagates towards the downstream side (see dashed arrows) [14]. The delay line method is used to determine the position of the hot spot on the meanderline by analyzing the arrival timestamps of the signals at the electrodes.

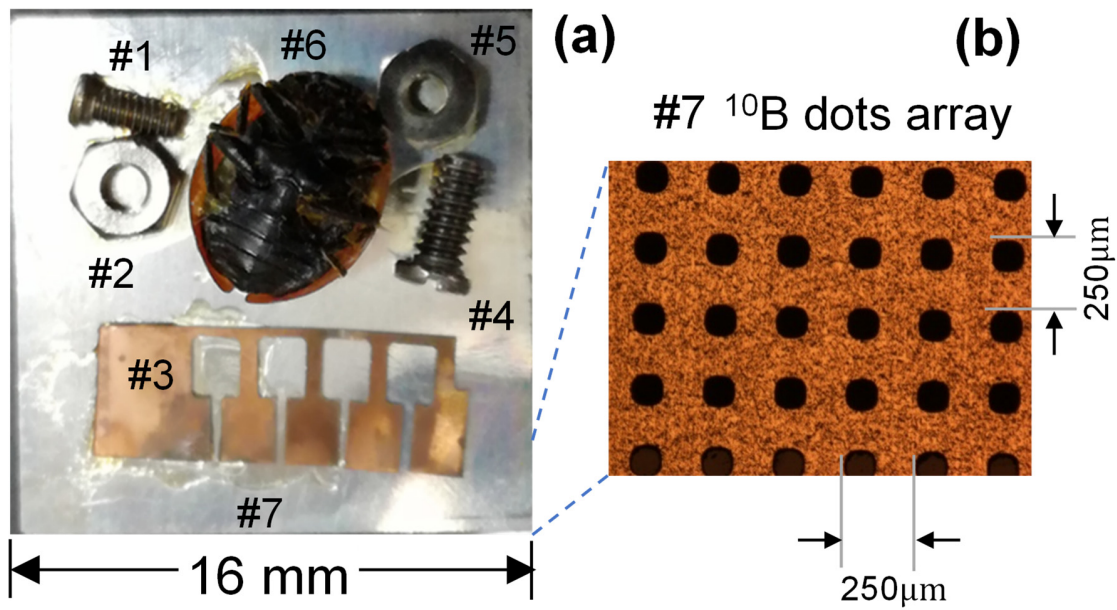

**Figure 2. (a)** Photograph of the test samples: a stainless-steel screw (sample #1), two Ni-plated brass nuts (sample #2, sample #5), a 0.1 mm-thick phosphor bronze plate (sample #3), a titanium screw (sample #4), and a seven-spotted ladybug (insect sample #6) are mounted on an Al plate. Under the Al plate, a 50 μm-thick stainless-steel mesh containing $^{10}$B dots acted as a neutron absorbing test pattern (sample #7). **(b)** Photograph of the test pattern of $^{10}$B dots, where each hole in the stainless-steel mesh is tightly filled with fine $^{10}$B particles. The mesh was designed to produce 100 μm holes with a pitch of 250 μm, but the holes are actually 106 μm due to the wet etching fabrication process. The edges of the holes in the metal are rounded due to this fabrication process. The pitch of the $^{10}$B dots is a suitable reference length for determining the characteristics of the imaging system.

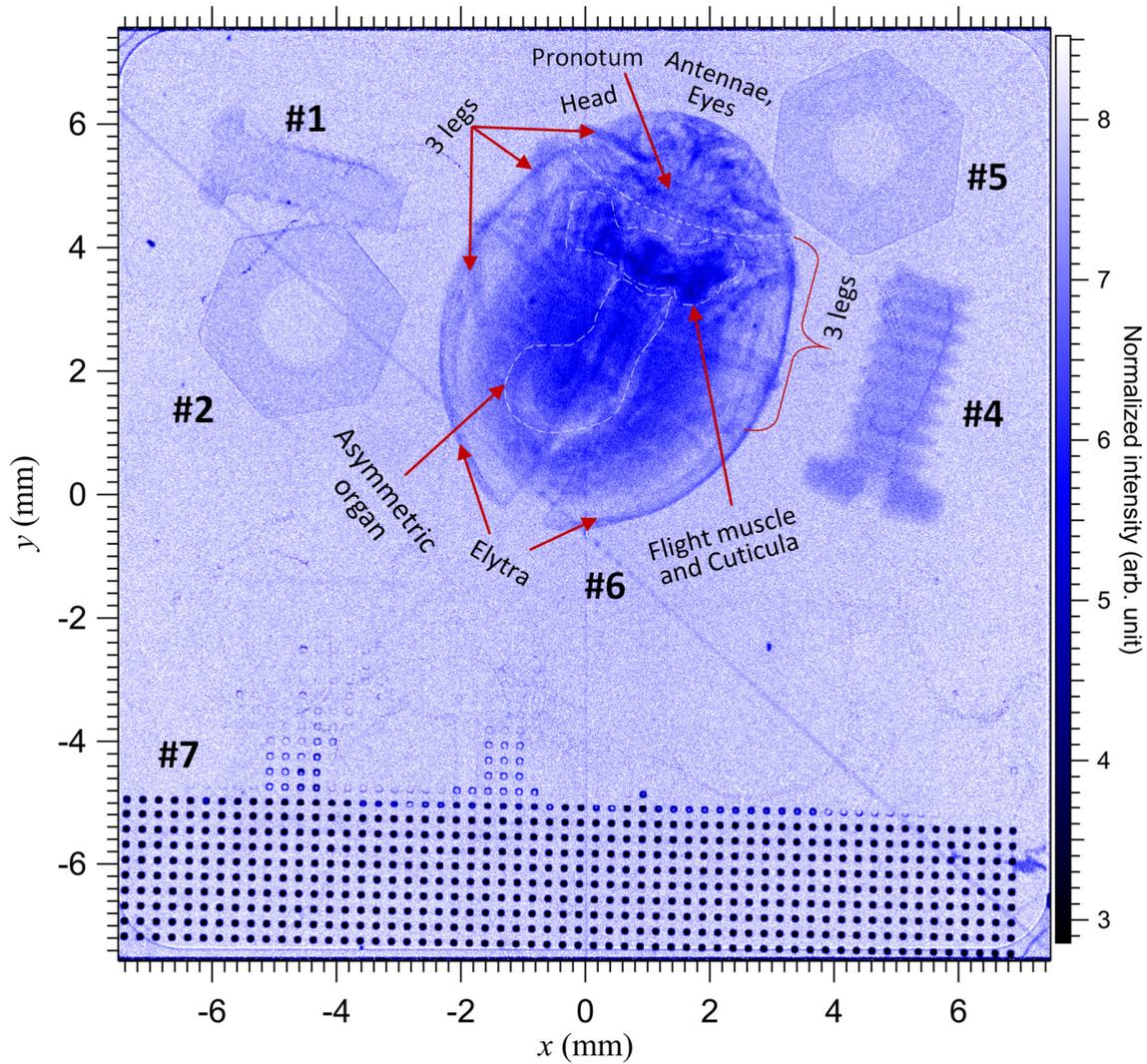

**Figure 3.** Transmission neutron image taken using CB-KID of the test samples with all neutron wavelengths (0.002 nm to 1.13 nm), where the intensity of each pixel is normalized by an image composed using short wavelength neutrons (0.002 nm to 0.052 nm). Average pixel size is $(5\Delta x_p, 5\Delta y_p)$ = (15.5 μm, 14.0 μm). For comparison with Fig. 2, we labeled the samples with #1, #2, #4, #5, #6, and #7. The intensity per pixel is shown with a relative color scale. The flight muscle and cuticula which support the flying wings and other organs are highlighted on the seven-spotted ladybug.

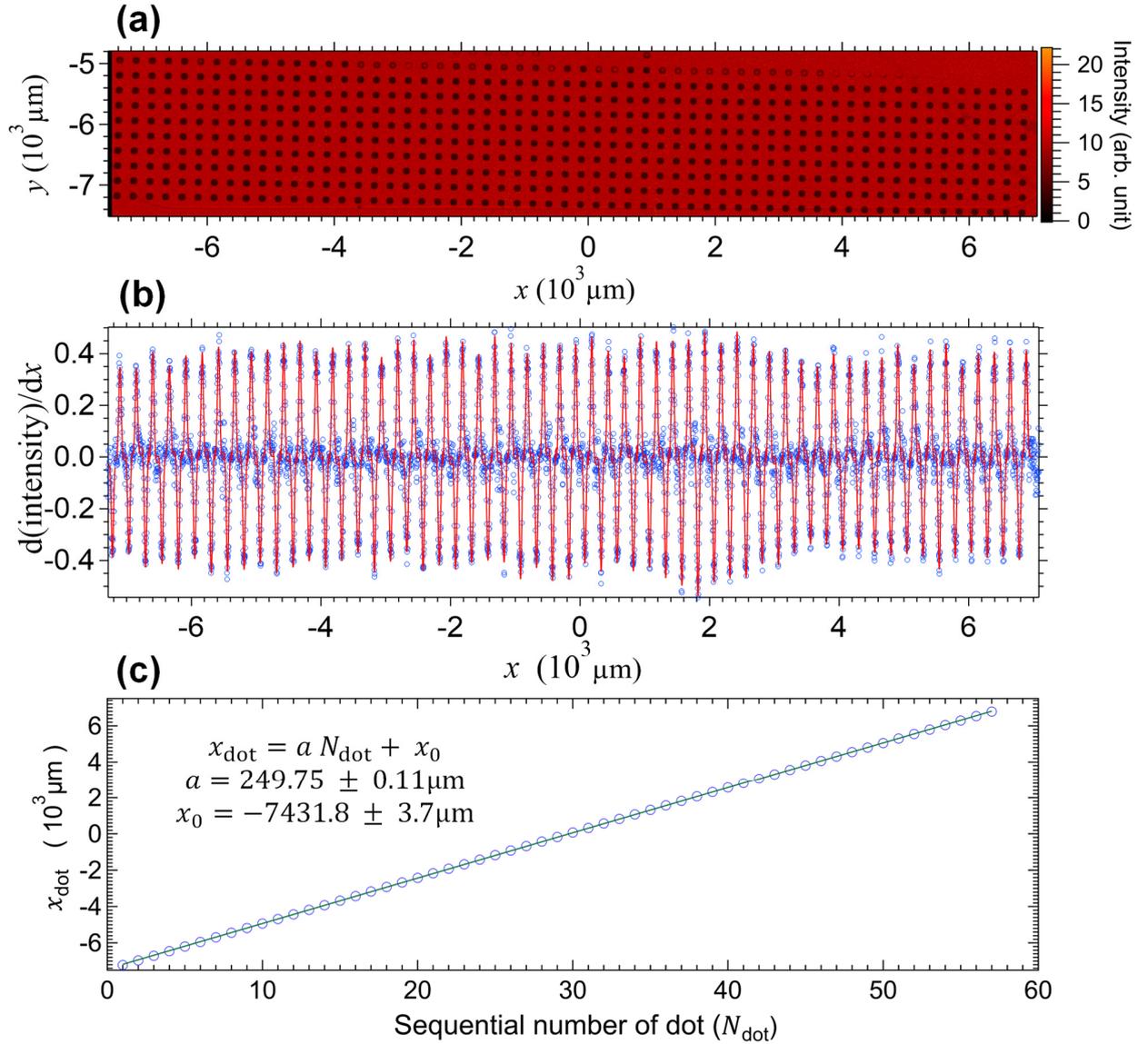

**Figure 4. (a)** Full-wavelength neutron transmission image of $^{10}$B dot array with an average pixel size of $(\Delta x_p, \Delta y_p) =$ (3.1 μm, 2.8 μm) normalized by the image taken with neutron wavelengths 0.002 nm to 0.052 nm. The number of pixels along the $X$ direction is ~4600 in this image. **(b)** Differential curves (blue circles) of the central-line profile of the $^{10}$B dots along the $X$ direction were obtained with an average pixel size of $(\Delta x_p, 5\Delta y_p) = $ (3.1 μm, 14.0 μm) for the first row of dots from the bottom of the sample, and were fitted by the least-squares method using Gaussian functions (red curves) to find the positions of the dot edges. We estimated the pitch of $^{10}$B dots by the distance between either positive neighboring peaks or negative neighboring peaks in the red curves fitted to the differential line profiles. **(c)** Positions of positive peaks $x_{\text{dot}}$ as a function of sequential number $N_{\text{dot}}$ of $^{10}$B dot for the first row from the bottom. Data points are fitted by a linear line $x_{\text{dot}} = a\, N_{\text{dot}} + x_0$, where $a$ is a gradient per dot.

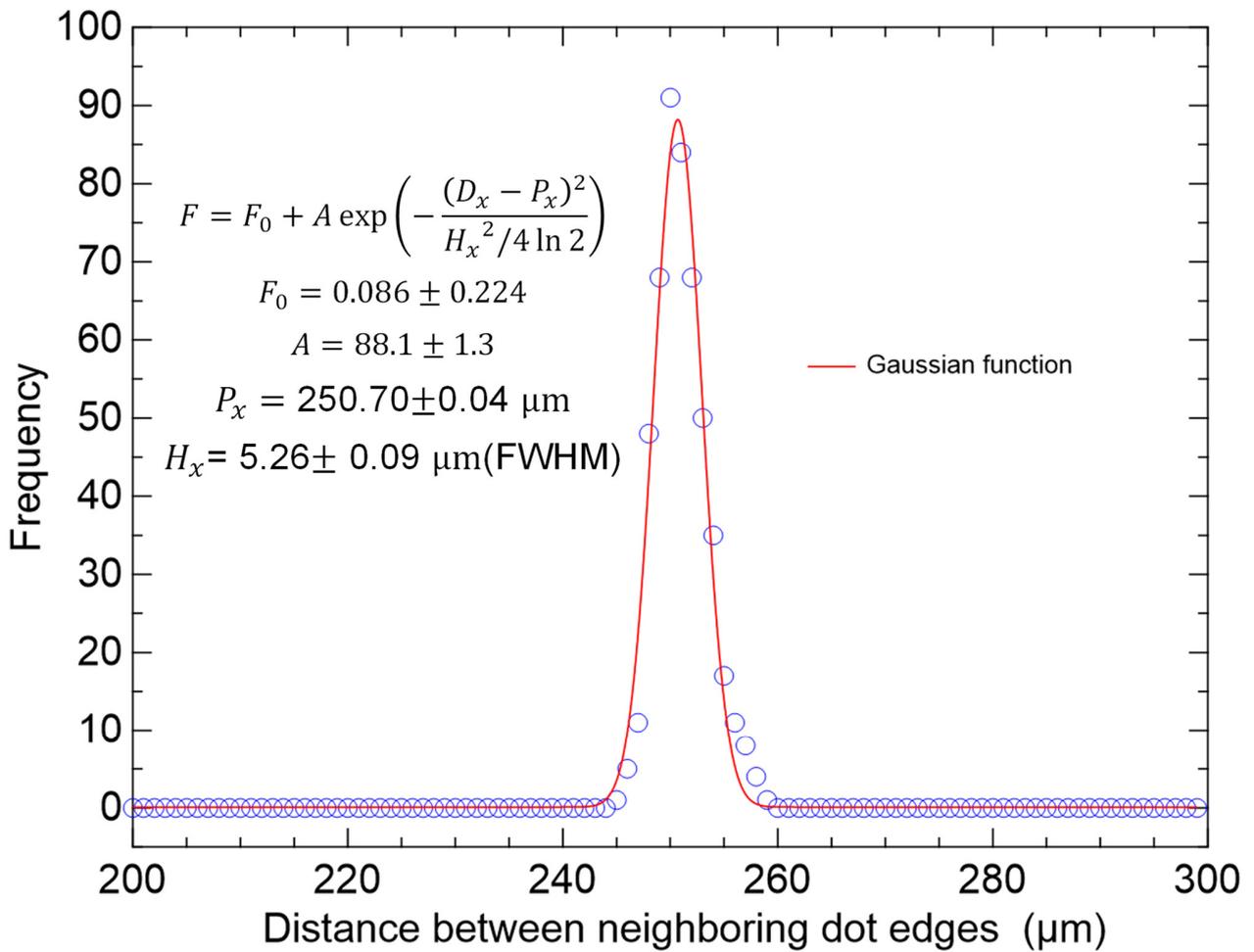

**Figure 5.** A typical distribution of the distances obtained between neighboring dots from the negative-peak positions along the $X$ direction. Data points are fitted by a Gaussian function (see red solid line). The central position $P_x$ of the Gaussian peak gives a pitch of 250.70 μm for the $^{10}$B dots, and this is in good agreement with the design value of 250 μm. The full width at half maximum $H_x$ of 5.26 μm shows a good homogeneity for pitches along the $X$ direction.

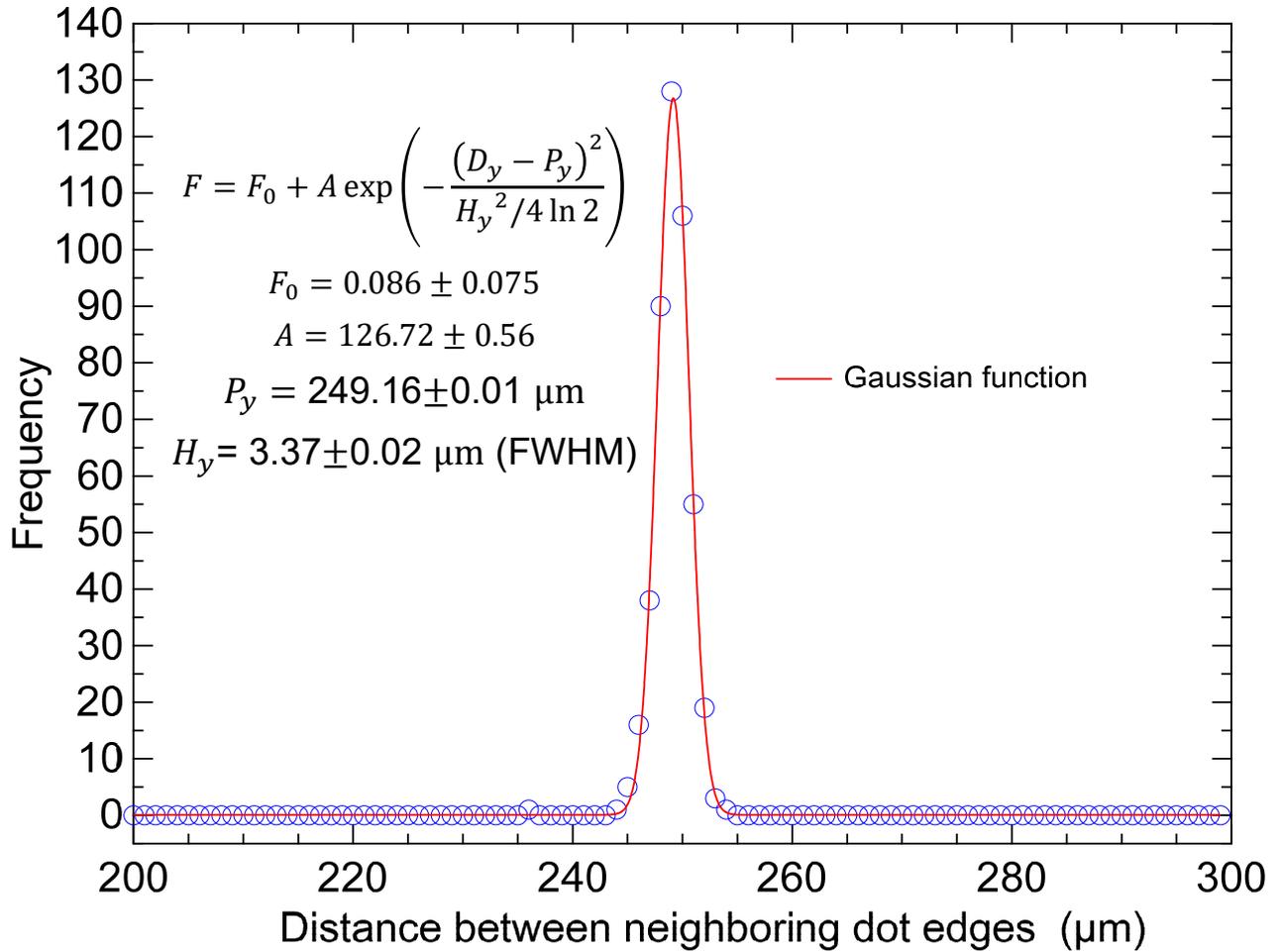

**Figure 6.** A typical distribution of the distances obtained between neighboring dots from the negative-peak positions along the *Y* direction. Data points are fitted by a Gaussian function (see red solid line). The central position $P_y$ of the Gaussian peak gives a pitch of 249.16 μm for the $^{10}$B dots, and this is in good agreement with the design value of 250 μm. The full width at half maximum $H_y$ was 3.37 μm for pitches along the *Y* direction.

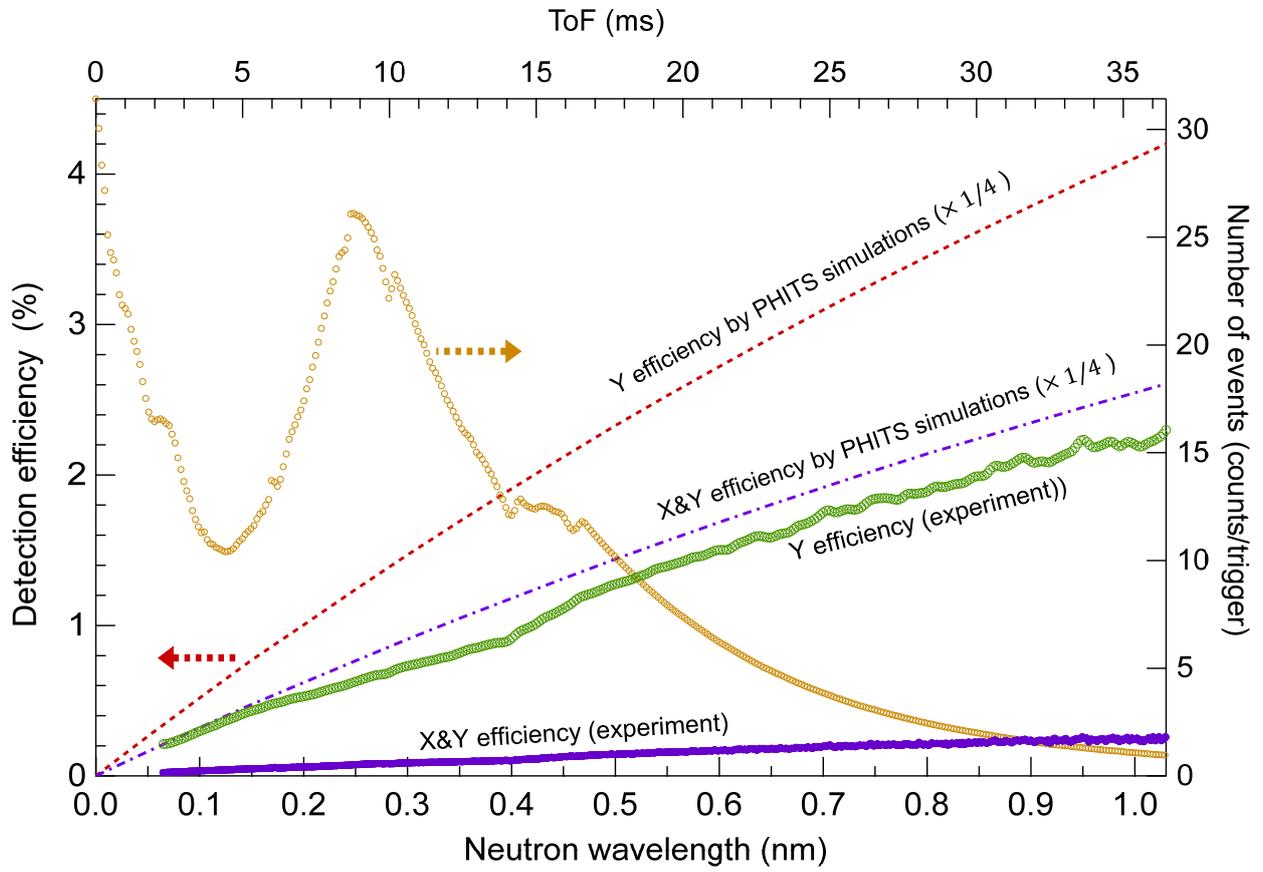

**Figure 7.** Experimental detection efficiency of a single meanderline (Y) and the simultaneous-detection efficiency of both the X and Y meanderlines as a function of neutron wavelength with 4 K detector temperature. The number of neutron events per trigger (40 ms) of pulsed neutrons obtained by the Y detector is shown as a function of wavelength, where the time bin was 0.1 ms and $A_d$ = 15 mm×15 mm. For comparison, detection efficiencies calculated from PHITS simulations (scaled by a factor of 1/4) are shown as a function of wavelength for the Y meanderline and for simultaneous detections of the same event by both the X and Y meanderlines [22]. A horizontal scale is also shown for time of flight (ToF) for the 14 m flight path of pulsed neutrons (top axis).